\documentclass[prl,twocolumn,showpacs,superscriptaddress,floatfix]{revtex4}

\usepackage{graphicx}% Include figure files
\usepackage{dcolumn} % Align table columns on decimal point
\usepackage{amsmath, amsthm, amssymb, amsfonts, bm}

\begin{document}
\title{Sculpting the Vortex State of a Spinor BEC}

\author{K. C. Wright}
\affiliation{Department of Physics and Astronomy, University of Rochester, Rochester, NY 14627}

\author{L. S. Leslie}
\affiliation{Institute of Optics and Laboratory for Laser Energetics, University of Rochester, Rochester, NY 14627}
         
\author{A. Hansen}
\affiliation{Department of Physics and Astronomy, University of Rochester, Rochester, NY 14627}

\author{N. P. Bigelow}
\affiliation{Department of Physics and Astronomy, University of Rochester, Rochester, NY 14627}
\affiliation{Institute of Optics and Laboratory for Laser Energetics, University of Rochester, Rochester, NY 14627}

\date{\today}  

\begin{abstract}
We use Raman-detuned laser pulses to achieve spatially-varying control of the amplitude and phase of the spinor order parameter of a Bose-Einstein condensate.  We present experimental results confirming precise radial and azimuthal control of amplitude and phase during the creation of vortex-antivortex superposition states.  
\end{abstract}

% 37.10.Vz 	Mechanical effects of light on atoms, molecules, and ions
% 03.75.Mn 	Multicomponent condensates; spinor condensates
% 03.75.Lm 	Tunneling, Josephson effect, Bose–Einstein condensates in periodic potentials, solitons, vortices, and topological excitations
% 37.25.+k 	Atom interferometry techniques
% 03.67.-a 	Quantum information
\maketitle

Many interesting properties of quantum spin fluids arise from the existence of topological structures in the order parameter. Vortices are among the most fundamental of these topological structures, and play an important role in the physics of superfluid helium, superconductors and neutron stars. Bose-Einstein condensates (BEC) of alkali metal atoms provide a unique setting in which to study superfluid vortices, and have enabled great progress in understanding these states \cite{Composite1}. 

Alkali metal BECs have multiple ground state sublevels, and are described by a multi-component order parameter. They can carry angular momentum in single-component vortices, or in multi-component structures such as coreless vortices \cite{HoSpinorPRL98,OhmiJPSJ98} and skyrmions \cite{AlKhawajaSkyrmionsN01,MuellerSpinPRA04}. Creating these structures requires precise control of the amplitudes and phases of the components of the order parameter.  In this Letter, we demonstrate unprecedented spatial control of a spinor BEC, using multi-pulse, multi-mode stimulated Raman interactions. These results show that it is now possible to deterministically create and study families of spin textures which have been predicted theoretically, but have yet to be directly observed \cite{IsoshimaQuantumJPSJ01,MizushimaAxisymmetricPRA02}. 

\begin{figure}[ht]
	\includegraphics[scale = 1.0]{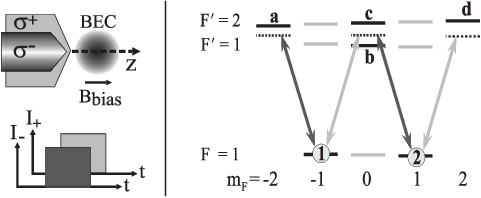}
	\caption{Experimental configuration.  A $^{87}$Rb BEC in the $\left|F=1,m_F=-1\right\rangle$ state is acted on by simultaneous, copropagating pulses of $\sigma^+$ (light gray) and $\sigma^-$ (dark gray) polarized laser light, with intensities $I_+$, $I_-$.   The fields are near resonant with the $5^2P_{1/2}$ excited states, and couple the states indicated in the energy level diagram (right).}
	\label{fig:levels}
\end{figure}

Our basic experimental configuration has been decribed previously \cite{WrightOpticalPRA08}. For simplicity, the results presented in this Letter focus on a specific interaction configuration using the $F=1$ ground state manifold of $^{87}$Rb, however, we have obtained similar results with the $F=2$ manifold. Essential details of the interaction configuration are depicted in Fig.~\ref{fig:levels}.  An untrapped BEC is acted on by $\sigma^+$ and $\sigma^-$ polarized laser beams which propagate collinearly, parallel to a small magnetic bias field ($\approx$ 1 Gauss). The beams are detuned near resonance with the $5^2P_{1/2}$ excited states ($D_1$ transitions), and satisfy the two-photon resonance condition for the Raman transition between the $\left|1\right\rangle \equiv \left|F=1,m_F=-1\right\rangle$ and $\left|2\right\rangle\equiv \left|F=1,m_F=1\right\rangle$ ground state sublevels. The individual excited state sublevels which interact with the fields are indicated in the diagram by the labels ($a$-$d$). 

\begin{figure}[ht]
	\includegraphics[scale = 1.0]{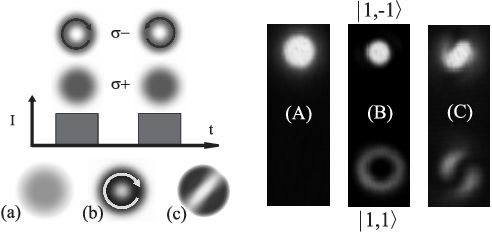}
	\caption{Vortex coupling and interference using beams of different orbital angular momentum.  Left: the laser pulse sequence and beam modes used to interact with the BEC, and the state of the BEC as the sequence progresses (a-c).  An initial laser pulse pair using Gaussian (G) and Laguerre-Gaussian (LG$_0^1$) beam modes causes the initally spin-polarized BEC ($a$) to evolve into a coreless vortex state ($b$).  Application of another pulse pair using G, LG$_0^{-1}$ beam modes, transforms this into a coreless vortex-antivortex superposition state ($c$). Right: Experimental images of the BEC ($A$-$C$) corresponding to schematics ($a$-$c$). The spin components are separated by a magnetic field gradient prior to absorption imaging.}
	\label{fig:setup}
\end{figure}

Complex spin textures can be created by applying spatially inhomogeneous optical fields to a spinor BEC in this configuration. Fig.~\ref{fig:setup} shows a procedure for creating first a coreless vortex, and then a vortex superposition state by using Laguerre-Gaussian (LG) optical vortex beams.  The coreless vortex is created by applying laser fields to the BEC, with the $\sigma^-$ polarized beam in an LG$_0^1$ spatial mode, and the $\sigma^+$ polarized beam in a Gaussian (G) spatial mode.  The difference in the optical field modes results in a transfer of orbital angular momentum (OAM) to a ring-shaped region of the BEC, which is coupled from the $\left|1,-1\right\rangle$ state to the $\left|1,1\right\rangle$ state. The center of this rotating ring in the $\left|1,1\right\rangle$ state is filled by atoms remaining in $\left|1,-1\right\rangle$; together these constitute a coreless vortex state. If a second laser pulse pair is then immediately applied which is identical to the first, except for a reversal in handedness of the $\sigma^-$ beam mode (to $LG_0^{-1}$), coupling occurs between this coreless vortex state and one of opposite sign.  

The final result of this two-stage procedure is a coherent superposition of a vortex and anti-vortex \cite{AndersenQuantizedPRL06,WrightOpticalPRA08} in the $\left|1,1\right\rangle$ component of the BEC.  When the density is measured these interfere to give the pattern seen in the schematic (c) and the experimental image (C).  The experimental results ($A$-$C$) are obtained by absorption imaging along the axis of the BEC after the components have been separated by a transverse magnetic field gradient. It should be emphasized that without this magnetic, Stern-Gerlach separation, the different components remain spatially overlapped as indicated by (a-c).  The interference pattern visible in (C) can be used to confirm the winding number of the vortex states, and is also an extremely sensitive indicator of spatial inhomogeneities in the evolution of the system.  This feature is of central importance to the results that follow.

\begin{figure}[ht]
	\includegraphics[scale = 1]{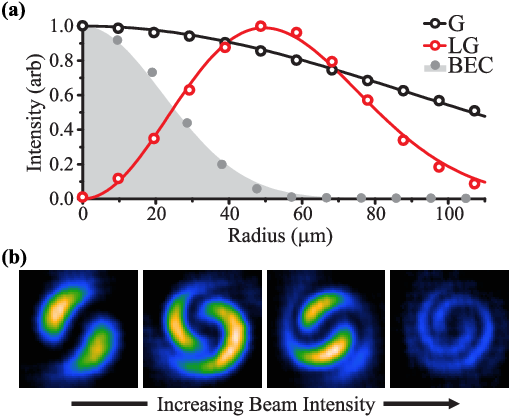}
	\caption{($a$) Radial intensity profiles for the G, LG beam modes used in the experiment, with the BEC density shown for comparison. The curves are Gaussian and LG mode functions fit to data points from azimuthally averaged lineouts of images. ($b$) Experimental images of vortex-antivortex superpositions created in $\left|1,1\right\rangle$ by the procedure in Fig.~\ref{fig:setup}, using the beam modes of ($a$).  The spiral distortion is caused by inhomogeneous light shifts, and increases with beam intensity. }
	\label{fig:inhomogeneous}
\end{figure}

One of the primary motivations of this work is to produce multicomponent coreless vortices in a BEC, with sufficient precision to study their dynamics and decay.  Some of the rich physics of multi-component vortices has been glimpsed in previous BEC experiments \cite{Composite2}, but the varieties of states that could be created were limited.  Raman-coupling of a spinor BEC allows deterministic creation of a wide variety of topological states, particularly within the same hyperfine manifold.  Many of these families of states may be difficult or impossible to generate with previous experimental techniques.  For example, the structure in Fig~\ref{fig:setup}(B) corresponds to that of a half-quantum, or `Alice' vortex \cite{IsoshimaQuantumJPSJ01}. 

The subtlest challenge in using stimulated Raman interactions to control a spinor BEC is dealing with spatially nonuniform light shifts. These can cause spatially inhomogeneous evolution of the BEC, making state preparation and dynamical studies very difficult.  The spiral distortion of the interference pattern in Fig.~\ref{fig:setup}(C) is an example of the consequences of these light shifts, which arise because the intensity of the beams is not uniform over the interaction region. We show here how these effects arise, and how to control and compensate for them.

Fig.~\ref{fig:inhomogeneous}(a) shows the radial dependence of the intensities of the beam modes used in the experiment.  The Gaussian beam intensity can be made uniform over the cross section of the BEC by increasing its spot size, however, the intensity of the LG beam mode cannot be made uniform, because the field vanishes at the central phase singularity.   The position-dependent intensity of the beams causes both the two-photon Rabi frequency, and the dynamic Stark shifts affecting the states, to be position-dependent.  These light shifts are also state-dependent, and cause the phase of the BEC components to evolve at different rates at different locations, as if in the presence of an inhomogeneous effective magnetic field created by the light fields.  

The spiral distortion in Fig.~\ref{fig:setup}(C) is a symptom of radial gradients in the relative phase of the BEC components, caused by these so-called Zeeman light shifts.  Increasing the effective temporal pulse area of the Raman beams magnifies this effect, resulting in tighter spirals in the patterns shown in Fig.~\ref{fig:inhomogeneous}b.  Reversing the handedness of the applied LG modes also reverses the spirals seen in the interference pattern. Creating a well-formed coreless vortex under these circumstances requires the elimination of this radial phase gradient.

Quantitative prediction of the response of the system to the laser fields is essential to achieving the necessary experimental control.  The light shifts that arise during the interaction of laser fields with a multi-level atom can be complex, however, it is possible to construct a simple and accurate model of systems like the one used in these experiments. Subject to certain restrictions on detuning and intensity, one can write a Hamiltonian for the system in the interaction picture \cite{ShoreTheory90}, apply the rotating wave approximation, and adiabatically eliminate the excited states \cite{FewellAdiabaticOC05}. Applying this procedure to the system described by Fig.~\ref{fig:levels} results in the following effective two-level Hamiltonian \cite{WrightRaman}, which can be used to describe the evolution of the components of the BEC state vector in an infinitesimal columnar section of the interaction region, parallel to the axis of symmetry.
\begin{equation}\label{eq:AEHF=1c}
\bm H=\hbar
\left[ \begin{array}{ccc} 
\frac{\delta}{2}-\chi_1^a I_{-} - \chi_1^{bc} I_{+}
&  
\eta_{12}^{bc}\sqrt{I_-I_+}e^{-i\xi_{\pm}}            
\\ [+0.5em]
\eta_{12}^{bc}\sqrt{I_-I_+} e^{i\xi_{\pm}}
& 
-\frac{\delta}{2}-\chi_2^d I_{+} - \chi_2^{bc} I_{-}
\end{array}\right]
\end{equation} 
   
	In this expression, $I_+$ ($I_-$) is the intensity of the $\sigma^+$ ($\sigma^-$) polarized laser field. The two-photon detuning is $\delta\equiv(E_1-E_2)/\hbar-(\omega_+-\omega_-)$.  $E_n$ is the energy of state $n$; $\omega_+$ ($\omega_-$) represents the frequency of the $\sigma^+$ ($\sigma^-$) polarized laser field. The quantity $\xi_{\pm}$ is the relative phase between the laser fields in the rotating basis defined by the frequency difference $(\omega_+-\omega_-)$; it is unimportant for infinite plane waves, but must be included to describe the spatial structure of beam modes carrying different OAM.  The intensity and relative phase represent the state of the optical fields at a point in the plane orthogonal to the propagation axis.  The atomic external degrees of freedom are represented in this approximation by the phase of the spin components; the atomic density is assumed to be low enough to neglect mean-field effects.

  The diagonal terms in this Hamiltonian containing the parameters $\chi$ are the state-dependent light shifts with which we are concerned. The off-diagonal term containing $\eta$ is the effective two-photon Rabi frequency for the coupling between $\left|1\right\rangle$ and $\left|2\right\rangle$.  The sub(super)scripts denote connections to specific ground (excited) states indicated by labels in Figure \ref{fig:levels}.  The coefficients determining the magnitude and sign of the light shifts ($\chi$) and the two-photon Rabi frequency ($\eta$) depend on the various dipole matrix elements, and the detuning of the beams from the Zeeman-shifted bare atomic transition frequencies.

\begin{figure}[ht]
	\includegraphics[scale = 1.0]{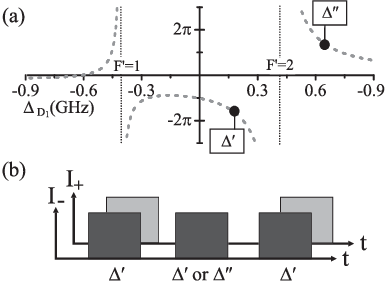}
	\caption{($a$) Calculated relative phase shift induced by a $\sigma^-$ polarized laser field as a function of detuning, for a pulse duration of 5 $\mu$s and fixed intensity of 44~mW/cm$^2$.  ($b$) Modified pulse sequence: A pulse of the $\sigma^-$ polarized field (LG mode) is inserted between the `transfer' pulses, causing an additional spatially inhomogeneous phase shift, with a detuning-dependent magnitude and sign. The detunings of the beams used in the experiment are indicated by the labels $\Delta'$ and $\Delta''$}
	\label{fig:phaseshift}
\end{figure}

The predictions of this model have allowed us to assert position-dependent control of the relative phase of the components of the BEC, and eliminate the distortions discussed above. This control is possible because the magnitude and sign of the Zeeman light shifts depend on the laser detuning. Fig.~\ref{fig:phaseshift}(a) shows the calculated relative phase shift caused by applying a short pulse of $\sigma^-$ polarized light to the BEC at different detunings, with the intensity and pulse duration held constant.  When the pulse is detuned between the $F'=1$ and $F'=2$ states, the energy of the $\left|1,-1\right\rangle$ state is decreased relative to that of the $\left|1,1\right\rangle$.  Detuning the field above the $F'=2$ excited state manifold causes the Zeeman light shift to reverse sign.  In the modified pulse sequence shown in Fig.~\ref{fig:phaseshift}(b), an additional pulse of the $\sigma^-$ polarized field is used to create an additional phase shift with identical position dependence to that nominally caused during the transfer pulses, but with a magnitude and sign that can be varied.  By choosing an appropriate detuning and intensity for this control pulse, the inhomogeneity in the system can be canceled out, leaving an undistorted coreless vortex state.  We can confirm this by observing the spiral-free interference pattern that would occur in the absence of the light shifts.  

 \begin{figure}[ht]
	\includegraphics[scale = 1]{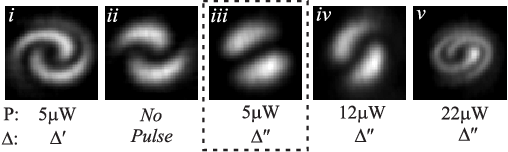}
	\caption{ Experimental images of vortex-antivortex superpositions in the $\left|1,1\right\rangle$ component of the BEC, created with the pulse sequence of Fig.~\ref{fig:phaseshift}(b).  ($i$) A control pulse with the same detuning as the transfer pulses tightens the spiral compared to the nominal result when no control pulse is applied ($ii$).  A control pulse detuned above the $F'=2$ manifold causes the light shifts in the sequence to cancel out, resulting in an undistorted interference pattern in the BEC ($iii$).  Increasing the intensity of the control pulse in the `cancelling' configuration overcompensates the radial phase gradients of the transfer pulses, creating spirals of opposite handedness ($iv$,$v$).}  
	\label{fig:phasecontrol}
\end{figure}

Fig.~\ref{fig:phasecontrol} shows the experimental results of implementing the pulse protocol of Fig.~\ref{fig:phaseshift}(b) to increase ($i$), reverse ($iv,v$), and remove ($iii$) the radial gradient in relative phase.  These are shown in comparison to the nominal result generated by the vortex interference pulse sequence without the extra control pulse ($ii$).  During the transfer pulses, both optical fields are detuned to $\Delta'$= 0.17~GHz, as indicated in Fig.~\ref{fig:phaseshift}($a$). For the result in Fig.~\ref{fig:phasecontrol}($i$), the control pulse detuning is the same as the transfer pulse pairs, resulting in an increased spiral in the same direction. For results ($iii$)-($v$), the detuning is set to $\Delta''$= 0.65~GHz, which results in a Zeeman light shift of opposite sign, with different magnitudes according to the pulse power used.   The key result in this set of images is ($iii$), in which the spatially inhomogeneous phase gradients created in the BEC during the transfer pulses have been canceled by the action of the control pulse. 

\begin{figure}[ht]
\centering
	\includegraphics[scale = 1.0]{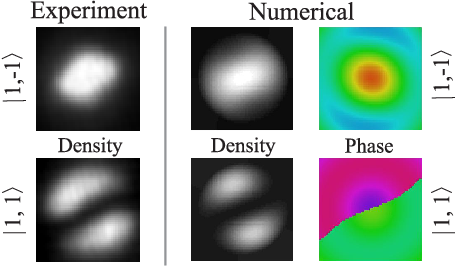}
	\caption{Comparison of experimental vortex interference patterns and corresponding numerical predictions from a model based on the Hamiltonian of Eq.~\ref{eq:AEHF=1c}.}
	\label{fig:tve}
\end{figure}

The images in Fig.~\ref{fig:tve} show the experimental results for the `canceling' configuration of Fig.~\ref{fig:phasecontrol}(iii), compared against the predictions of a numerical model based on Eq.~\ref{eq:AEHF=1c}.  There is close agreement between the experimental and numerical density distributions, which validates the approximations used in constructing the model.  The numerical result also provides information about the phase of the BEC that is difficult to access experimentally. Note that the calculated phase of the $\left|1,1\right\rangle$ state shows the expected $\pi$ discontinuity along the density node.  The slight depression in the phase of both states arises from the uncompensated scalar part of the light shifts, which does not affect the spin dynamics.   This model can be used as a guide for creating well-defined perturbations of a coreless vortex state, which may prove to be a useful feature for future dynamical studies. 

The results summarized in Figs. \ref{fig:phasecontrol} and \ref{fig:tve} demonstrate that it is possible to create complex structures such as the coreless vortex of Fig.~\ref{fig:setup}(b) in a well-controlled manner using Raman-coupling methods.  The dependence of the interaction on excited state structure and field detunings allows tremendous flexibility in tailoring the evolution of the system for specific experimental goals.  The success of the approach presented indicates that proposed schemes for the storage, manipulation, and retrieval of information in a spinor BEC \cite{KapaleVortexPRL05} may be feasible, even in the presence of complicated multilevel effects.  Furthermore, understanding of the principles illustrated in this Letter are crucial to creating and observing the evolution of coreless vortices, for example in an optical dipole trap. Further experiments demonstrating the creation of important classes of two-dimensional skyrmion states such as the Anderson-Toulouse \cite{AndersonPhasePRL77} and Mermin-Ho \cite{MerminCirculationPRL76} coreless vortices are in progress in both the $F=1$ and $F=2$ Zeeman manifolds.

This work was supported by the NSF and ARO. LSL is grateful for a Horton Fellowship.

\end{document}